\documentclass[aps,prl,twocolumn,floatfix]{revtex4-1}
\usepackage{times}
\usepackage{epsfig}
\usepackage{subfig}
\usepackage{color}
\usepackage{fancyhdr}
\usepackage{multirow}
\usepackage{array}

\usepackage{graphics,graphicx}
\usepackage{bm,bbm}
\usepackage{mathrsfs}
\usepackage{calc}
\usepackage{epsfig}
\usepackage{amssymb,amsmath,amsfonts}
\usepackage{multirow}
\usepackage{array}

\def\bpmat{\begin{pmatrix}}
\def\epmat{\end{pmatrix}}
\def\bmat{\begin{matrix}}
\def\emat{\end{matrix}}

\def\1{\mbox{1\hskip-.25em l}}

\def\bea{\begin{eqnarray}}
\def\eea{\end{eqnarray}}
\def\be{\begin{equation}}
\def\ee{\end{equation}}

\def\beq{\begin{equation}}
\def\eeq{\end{equation}}

\begin{document}

\title{The time-asymmetric quantum state exchange mechanism}

\author{ Ido Gilary$^{(1)}$, Alexei A. Mailybaev$^{(3)}$, and Nimrod Moiseyev$^{(1,2)}$}

\affiliation{$^{(1)}$Schulich Faculty of Chemistry and $^{(2)}$Faculty of Physics,
Technion-Israel Institute of Technology, Haifa, 32000, Israel, $^{(3)}$Instituto Nacional de Matem\'{a}tica Pura e Aplicada - IMPA,
Estrada Dona Castorina 110, 22460-320 Rio de Janeiro, RJ, Brazil}

\begin{abstract}
We show here that due to non-adiabatic couplings in decaying systems applying the same time-dependent protocol in the forward and reverse direction to the same mixed initial state leads to different final pure states.  In particular, in laser driven molecular systems applying a specifically chosen positively chirped laser pulse or an equivalent negatively chirped laser pulse yields entirely different final vibrational states. This phenomenon occurs when the laser frequency and intensity  are  slowly varied around an exceptional point (EP) in the laser intensity and frequency parameter space where
the non-hermitian spectrum of the problem is degenerate. The protocol implies that a positively chirped laser pulse traces a counter-clockwise loop in time in the laser parameters' space whereas a negatively chirped pulse follows the same loop in the clockwise direction. According to this protocol one can choose the final pure state from any initial state. The obtained results imply the intrinsic non-adiabaticity of quantum transport around an EP, and offer a way to observe the EP experimentally in time-dependent quantum systems.

\end{abstract}

\maketitle

For open quantum systems where the effective Hamiltonian is non-hermitian (NH) the non crossing rule \cite{non-crossing} is replaced by an intersection of  two
{complex} energy levels associated with two eigenfunctions of the NH Hamiltonian that have the same symmetry.
Let us consider $2\times2$ Hamiltonian matrix $H$ which depends on potential parameters $q_1$ and $q_2$. These can be for instance the laser frequency and intensity when light interacts with two normal modes of a molecule.

In open quantum system where the effective Hamiltonian is NH all matrix elements can  attain  complex values. The complex diagonal matrix elements are associated with meta-stable (resonance) states, such that $-2Im H_{11}$ and $-2Im H_{22}$ are the decay rates of the meta-stable states.
The eigenvalues of this NH Hamiltonian are degenerate when
$\Delta=(H_{11}-H_{22})^2+4H_{12}H_{21}=0$ even though all matrix elements
are different from zero. This situation is very different from the hermitian (standard) case where crossing requires $H_{12}=H_{21}=0$ and $H_{11}=H_{22}$. At the crossing point a non-hermitian degeneracy (NHD) is obtained when the following two equations are satisfied:
\begin{eqnarray}
\label{NHD1}
&&Re[H_{11}-H_{22}]=\mp 2Im [H_{12}H_{21}]^{1/2}\\
&&Im[H_{11}-H_{22}]=\pm 2Re[H_{12}H_{21}]^{1/2}.
\label{NHD2}
\end{eqnarray}

NHD is very different in its nature from hermitian degeneracy. NHD is obtained at the crossing point denoted by $(q_1^{EP},q_2^{EP})$, where the
two eigenvalues coalesce and form a branch point (BP) in the complex energy spectrum. At the BP the first order derivatives of the eigenvalues with respect to $q_1$ or $q_2$ acquire infinitely large values (see for example Chapter 9 in Ref.\cite{NHQM-BOOK}). This BP is also known  as an exceptional point (EP) in the energy spectrum\cite{KATO,Heiss}.
Moreover, at the BP (EP) not only the eigenvalues coalesce but also the corresponding eigenvectors. Such a phenomenon can never occur in standard QM. In NHQM as $q_1\to q_1^{EP}$ and  $q_2\to q_2^{EP}$ the two bi-orthogonal eigenvectors of the complex non-hermitian Hamiltonian matrix coalesce. Rather than two different bi-orthogonal eigenvectors we get only one eigenvector which
is self-orthogonal (with respect to the c-product)\cite{CS-REVIEW1,cproduct}. As proved in Refs.\cite{SO,Arnold}, the NHD (i.e., EP)
is a typical phenomenon in NHQM.

Since there is no analog to this situation in hermitian QM it was believed for many years that the EP is a mathematical object only as it appeared mostly in complex scaled Hamiltonians \cite{SO}. Yet, even as a mathematical object it was found to be a useful concept that helps to explain experimental results which could not be explained otherwise (see Ref.\cite{EP4} where the mysterious sharp peaks in the cross section  measurements of electron scattering from hydrogen molecule were explained for the first time).  In the last decade it became clear that EPs are not only mathematical objects but play a major role also in actual measurable phenomena. Different manifestations of the EPs
have been described in optics\cite{EP1},
in superconductors\cite{EP5},
 in quantum phase transitions in a system of
interacting bosons\cite{EP6}, in electric field oscillations in
microwave cavities\cite{EP7}, and in PT-symmetric waveguides\cite{EP8}.
  So far there are \underline{no} experimental results regarding EPs in atomic, molecular, or biophysical systems.

Before proceeding we should mention the most striking phenomenon induced by  EPs which has no equivalent in hermitian QM: {the state exchange phenomenon}.
The state exchange phenomenon can be illustrated as follows:
Consider an arbitrarily variation of the two parameters $(q_1,q_2)$ which depend on angle variable $\varphi$ in a closed loop around an EP $(q_1^{EP},q_2^{EP})$. The two instantaneous eigenvalues
 are given by,
\begin{equation}
  E_\pm(\varphi)= \frac{H_{11}+H_{22}\pm \sqrt {\Delta (\varphi)}}{2}
\end{equation}
where the quantity $\Delta(\varphi)$ makes a circle around the origin in complex plane with a change of $\varphi$. Thus, it is easy to see that
$E_\pm(0)=E_\mp(2\pi)$.
Instead of the Berry phase
   which is obtained when cycling around a conical intersection, when cycling
   around an EP  one state flips into the other (see Chapter 9 in
Ref.\cite{NHQM-BOOK}).
To the best of our knowledge the only measurement of this state exchange  phenomenon was carried out by Richter and his co-workers  in microwave experiments\cite{EP7}. The association of the
state exchange  phenomenon with {molecular} system
 was made by Lefebvre and his co-workers\cite{ROLAND}.

The model $2\times2$ Hamiltonian matrix which was discussed above  can describe in the NH case two coupled resonance states, where $H_{11}$ is the complex energy of the atomic, molecular or mesoscopic  resonance state that absorbed one photon, $H_{11}=E_1-i\Gamma_1+\hbar\omega$, while $H_{22}=E_2-i\Gamma_2$ is the complex energy of the excited resonance state. $\Gamma_1>0$ and $\Gamma_2>0$ are the decay rates of the two resonance states. The coupling term as usual is given by $H_{12}=H_{21}=\epsilon_0 d_{12}/2$ where $\epsilon_0$ is the maximum laser field amplitude and $d_{12}$ is the complex dipole transition matrix element. The NH Hamiltonian matrix can be now rewritten such that

\begin{equation} \label{MATRIX2} H= \left(\begin{array}{cc} E_1+\hbar\omega +\frac{i\Delta\Gamma}{2} & \frac{\epsilon_0  d_{12}}{2} \\ \frac{\epsilon_0 d_{12}}{2} & E_2 -\frac{i\Delta\Gamma}{2}  \end{array}\right) -i\frac{\Gamma_1+\Gamma_2}{2}
\left(\begin{array}{cc} 1 & 0 \\ 0 & 1 \end{array} \right).
\end{equation}

where $\Delta\Gamma= \Gamma_2-\Gamma_1$. As one can see from Eq.\ref{MATRIX2} relative gain and loss states are obtained (e.g., when $\Delta\Gamma>0$ then one state has a relative gain while the other resonance state has a relative loss).
EP is obtained when Eqs. \ref{NHD1},\ref{NHD2} are satisfied. Consequently, an EP in the spectrum  is obtained when the maximum field amplitude is  $\epsilon_0^{EP}=\Delta\Gamma/Re[d_{12}]$
and the laser frequency is equal to
$\omega^{EP}=(E_2-E_1-Im[d_{12}]\epsilon_0^{EP})/\hbar$.
When the laser field is strong enough to allow a multi-photon absorption the calculations of the conditions for EP are slightly more complicated but achievable.

In closed systems adiabatic solutions converge to the exact solutions of the TDSE in the limit of infinitely slow variation of the potential parameters. In open systems the situation is very different. In contrast to hermitian systems, in open systems for almost any path in parameter space the non-adiabatic couplings become {more} significant the slower the potential parameters are varied. As a result the adiabatic theorem often breaks down in open quantum systems. Let us explain this for our $2\times2$  model hamiltonian when the two potential parameters $q_1$ and $q_2$  are time dependent parameters. These can be external field parameters such as laser frequency and intensity. The conventional adiabatic approximation is associated with the eigenvalues $E^{ad}_{\pm}(q_1, q_2)$ and eigenfunctions $\phi_\pm^{ad}(q_1,q_2)$ of the Hamiltonian matrix in Eq.\ref{MATRIX2}.
The dynamical non-adiabatic corrections to the solutions of the TDSE result from the dependence of the potential parameters $q_1$ and $q_2$ on time. The dynamical non-adiabatic hamiltonian matrix elements couple different adiabatic states and are given by the  following matrix elements,
\begin{eqnarray}
\label{NA}
&&V_{+/-}^{NA}= {\cal V}_{+/-}e^{+i\oint_{0}^{T} \Delta E^{ad}(q_1(t),q_2(t))dt }  \nonumber \\
&&V_{-/+}^{NA}= {\cal V}_{-/+}e^{-i\oint_{0}^{T}\Delta E^{ad}(q_1(t),q_2(t))dt }
\end{eqnarray}
where:
\begin{eqnarray}
\label{NA1}
&& {\cal V}_{+/-}=\langle \phi_+^{ad}(q_1,q_2)|\dot{q}_1\frac{\partial}{\partial q}_1+\dot{q}_2\frac{\partial}{\partial q}_2|\phi_-^{ad}(q_1,q_2)\rangle \nonumber \\
&& {\cal V}_{-/+}= \langle \phi_-^{ad}(q_1,q_2)|\dot{q}_1\frac{\partial}{\partial q}_1+\dot{q}_2\frac{\partial}{\partial q}_2|\phi_+^{ad}(q_1,q_2)\rangle
\end{eqnarray}
and,
\begin{eqnarray}
\label{NA1last}
&&\Delta E^{ad}(q_1,q_2)= E_+^{ad}(q_1,q_2)- E_-^{ad}(q_1,q_2)\nonumber \\ &&
\equiv \Delta{\cal E}^{ad}(q_1,q_2)-i\Delta\Gamma^{ad}(q_1,q_2).
\end{eqnarray}
Here $T$ is the duration of the loop in the parameter space.
For closed systems where the Hamiltonian is hermitian  $\Delta E^{ad}(q_1,q_2)$ has real values only. Therefore the exponents in Eq.\ref{NA} are just a phase factor, and the non-adiabatic corrections vanish as $T\to\infty$ (i.e., the variation of the potential parameters is arbitrarily slow).

In open system where we are dealing with resonances
the energy difference might be complex. The imaginary part of the energy difference,  $\Delta\Gamma^{ad}>0$  for instance, leads  to $V_{+/-}^{NA}\to \infty$ while  $V_{-/+}^{NA}\to 0$ as $T\to\infty$. The exponential divergence in $T$ of the exponent in $V_{+/-}^{NA}$ easily overcomes the $1/T$ suppression (responsible for the Hermitian adiabatic theorem) that is associated with the pre-exponential terms in Eq.\ref{NA} that contain the time derivatives of the potential parameters $\dot{q}_{1,2}$. This implies that in the limit of slow evolution only one state evolves adiabatically while the other state behaves  non-adiabatically. The adiabatic state is the one which decays slower. For other states the adiabatic solution is not valid even approximately, making the adiabatic flip often discussed in the literature \cite{ROLAND,MailybaevEtAl2005} impossible.

\begin{figure}[h]
\begin{center}
    \includegraphics[width=1\columnwidth, angle=0,scale=0.75,
draft=false,clip=true,keepaspectratio=true]{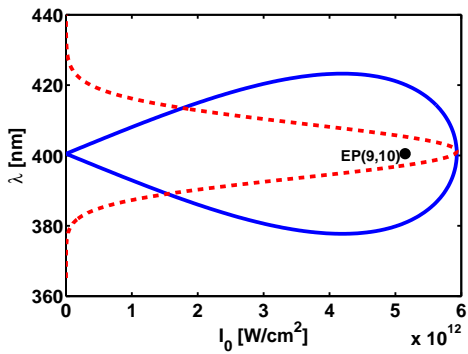}
\caption{
Two loops in the parameter space of the laser intensity and wavelength. When the laser intensity is turned on the $9th$ and the $10th$  vibrational states of the molecular ion become resonances.
The EP is obtained when the two resonances become degenerate  not-hermitian eigenstates. Both loops encircle the
EP, where the two states coalesce.
}
\label{Fig.1}
    \end{center}
\end{figure}

Let us assume that our protocol implies that the external potential parameters $(q_1,q_2)$
are varied in time in a closed loop which encircles the EP (see for example  Fig.\ref{Fig.1}).  This EP is obtained at the values of the external parameters $(q_1^{EP},q_2^{EP})$. Note that {this dynamical protocol requires to solve the time-dependent Scr\"odinger equation}.  The only adiabatic solution which describes correctly the dynamics is the longest-lived resonance state.
The lifetimes of the adiabatic resonance states are obtained by averaging the decay rates over the entire closed loop in the potential parameter space. That is, the inverse lifetime of the  adiabatic states denoted by $\pm$ are given by $T^{-1}\int_0^T \Gamma_{\pm}^{ad}dt$ where $\Gamma^{ad}_{\pm}$ is the decay rate at any given point on the closed loop.

A key point in the understanding the difference between the dynamics for bound and decaying systems is to realize how the non-adiabatic dynamical correction terms couple different adiabatic solutions. There is an asymmetric phenomenon in the calculation of the strength of the dynamical non-adiabatic coupling  between two adiabatic resonance  solutions which do not exist between two bound adiabatic solutions.  The strength of the non-adiabatic coupling term that induces a transition from the $|\phi_+^{ad}\rangle$ state to the $|\phi_-^{ad}\rangle$ state is  proportional to $e^{-\int_0^T \Delta\Gamma^{ad} dt}$ while the strength of the coupling that induces the  transition from the $|\phi_-^{ad}\rangle$ to the $|\phi_+^{ad}\rangle$ adiabatic resonance solution is the inverse of this expression, i.e.,
$e^{+\int_0^T \Delta\Gamma^{ad} dt}$ see Eq.\ref{NA}--\ref{NA1last}. {This asymmetric dynamical non-adiabatic effect stands behind our discovery that at the end of the propagation the system is found in the longest-lived pure state, irrespective of the initial condition. This effect is not due to the "evaporation" of the shorter-lived adiabatic states. It happens  because the short-lived adiabatic states are transformed into the long-lived adiabatic states during the time propagation process.}
This phenomenon  is first described in our paper published in Ref.\cite{RAAM}. In Ref.\cite{IDO-JPB} this exchange-state phenomenon was illustrated for a real physical  system when $H_2^+$ interacts with chirped laser pulses.

\begin{figure}[!h]
    \includegraphics[width=1\columnwidth, angle=0,scale=0.85,
draft=false,clip=true,keepaspectratio=true]{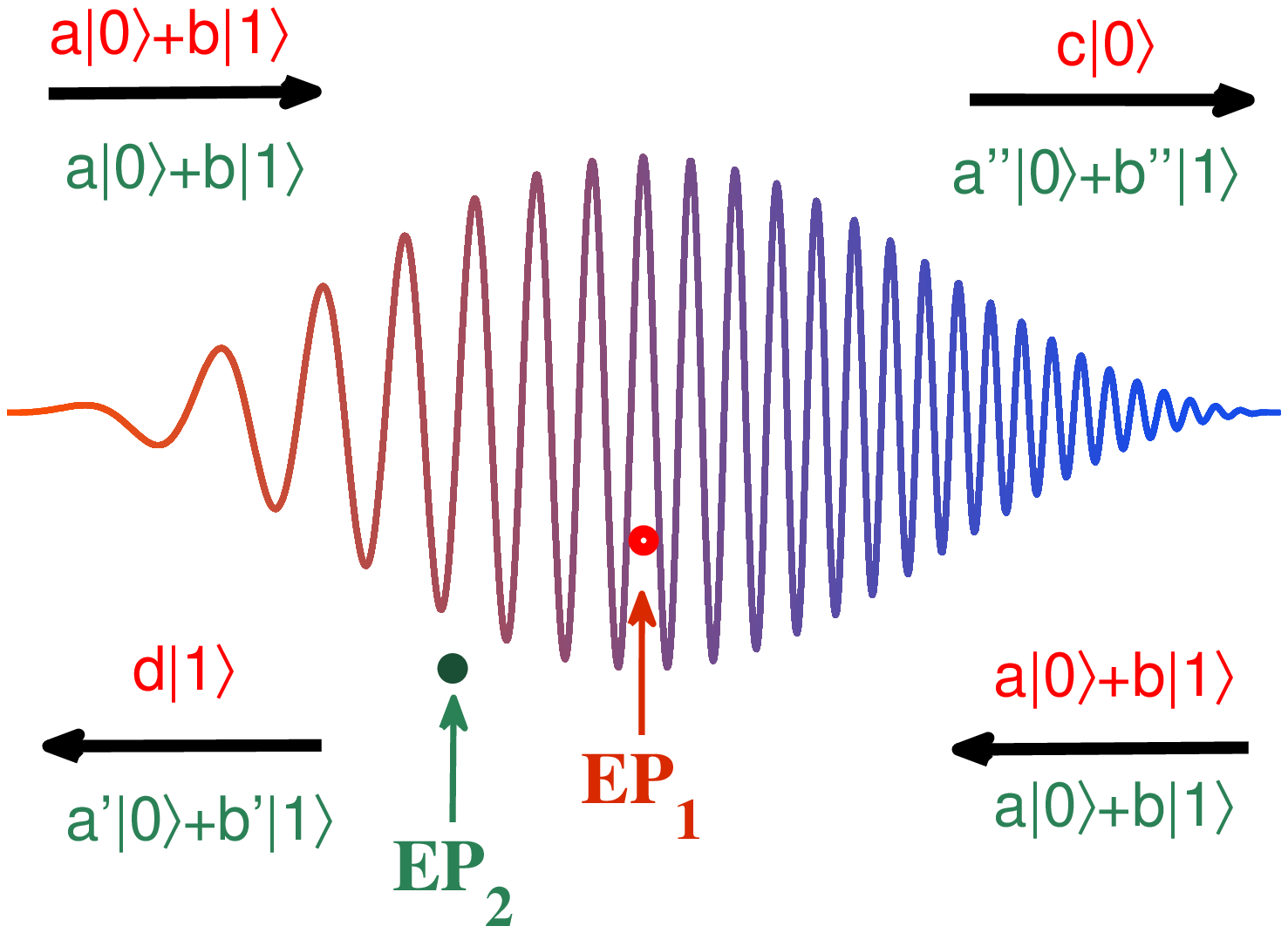}
\caption{A schematic representation of a diode-like device based on the NA-TAS
mechanism. The device can be an optical WG (passive diode-like) or an
electronic one related to any of the applications described below.\\
The multi-color line shows the external field as a function of the propagation in the device. This field is characterized by two parameters, the amplitude and frequency which both change during propagation.
Two options are presented. In red the parameters are varied in a closed loop encircling an EP (EP$_1$) whereas in green the EP is outside of the loop (EP$_2$).
The input in the two cases is the same, i.e. superposition of two modes.
The output for the two cases is very different. In red we
show that only when the EP is inside the loop of the time varying
parameters only ONE mode is obtained, $|0\rangle$ or $|1 \rangle$. The output depends on the directional propagation in the device showing its asymmetric behavior.}
\label{SCHEME}
\end{figure}

We are now at core of the universal asymmetric state exchange phenomenon
which is the focus of this study. It is based on the observation
 that the integral $\int_0^T \Delta\Gamma^{ad} dt$  changes sign when the loop in parameter space changes from the clockwise direction to the counter clockwise direction if and only if the closed loop encircles an EP (this is due to the exchange of the instantaneous states which is the property of the EP described above).
This means that the dynamical protocol described above imposes specific asymmetry for our time dependent hamiltonian.
The consequences are dramatic. If by applying the dynamical protocol in the clockwise direction the $|\phi^{ad}_+\rangle$  state is obtained as the external parameters return to their initial values (independently of the initial state). The $|\phi_-^{ad}\rangle$ state will be obtained when the same protocol is applied in the counter clockwise direction. This way we can control the dynamics and produce a quantum diode-like device for atomic, molecular or optical systems,
such that the output depends on the direction in which one enters the device. A general scheme for such device is depicted in Fig.\ref{SCHEME}.

\begin{figure}[h]
\begin{center}
\subfloat[][]{
\includegraphics[width=1\columnwidth, angle=0,scale=0.45,
draft=false,clip=true,keepaspectratio=true]{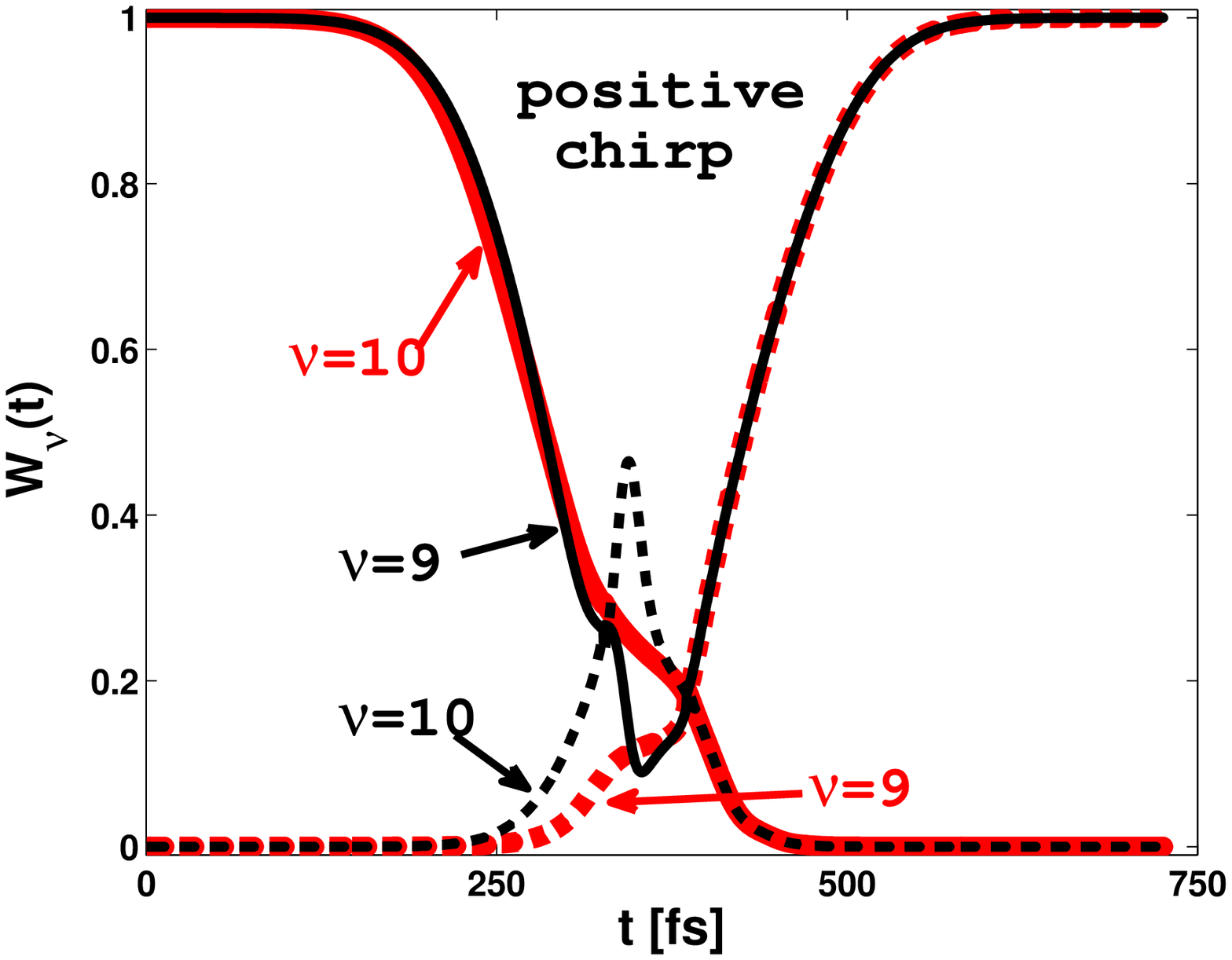}}
\quad
\subfloat[][]{  \includegraphics[width=1\columnwidth, angle=0,scale=0.45,
draft=false,clip=true,keepaspectratio=true]{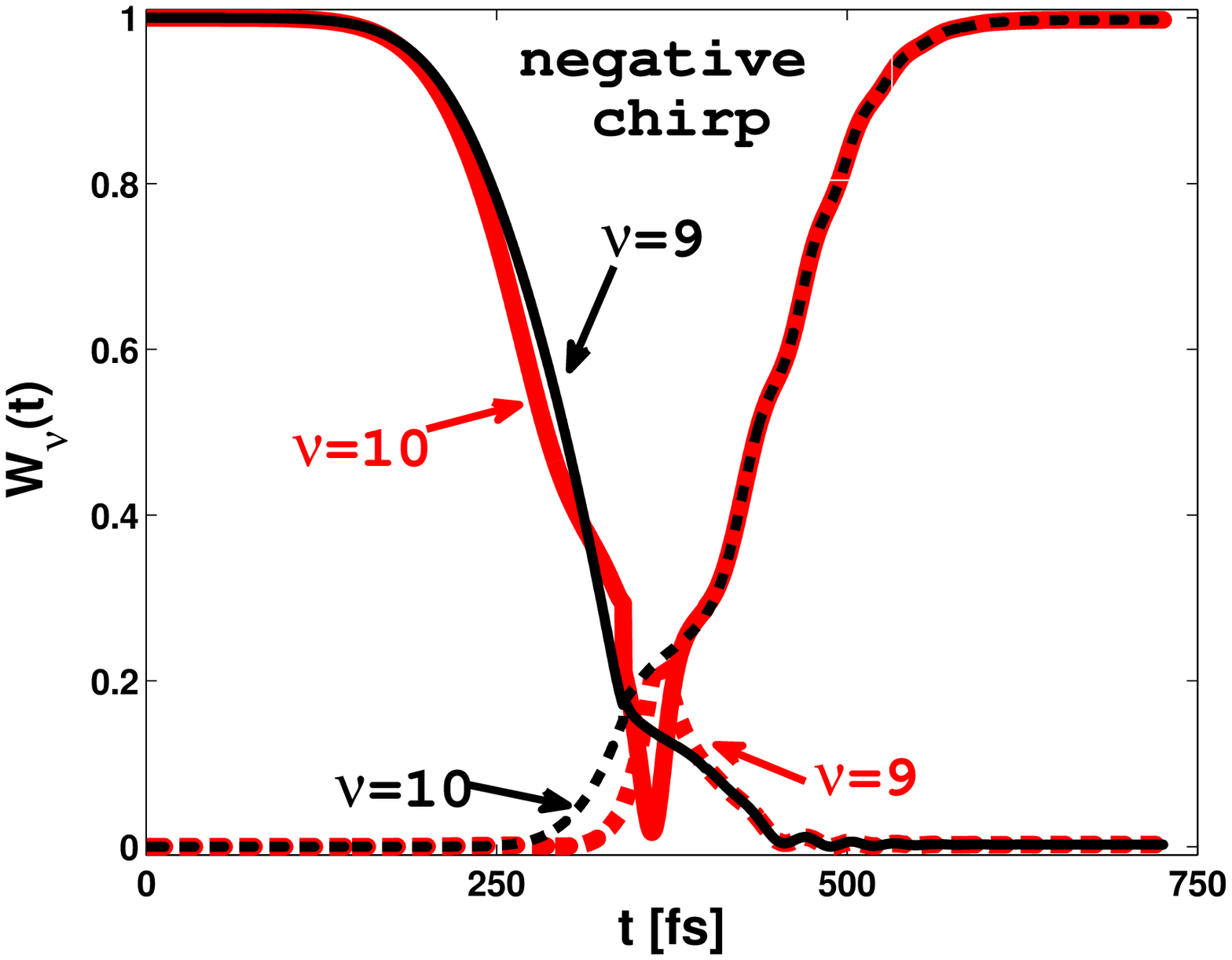}}
\caption{
The asymmetric state-exchange mechanism in H$_2^+$. $W_\nu(t)$ gives the projection of the propagated $H_2^+$ wavepacket (WP) during a chirped laser pulse on the field-free vibrational states of the molecular ions, $|\nu=9\rangle$ and $|\nu=10\rangle$. In this calculation we
normalized the propagated WP to unity at any given time in order to compensate for the decay. The thin black lines describe the solution which starts at the $|\nu=9\rangle$ vibrational state whereas the thick red lines describe the solution which starts at the $|\nu=10\rangle$ vibrational state. The chirp changes the laser wavelength and intensity in time according to the dashed red loop encircling the EP shown in Fig.~1. (a) A positive chirped laser pulse corresponds to a clockwise trajectory on the loop. (b) A negative chirped laser pulse corresponds to a counter-clockwise trajectory on the loop. Similar results are obtained when we follow a trajectory along the blue solid line in Fig.~\ref{Fig.1} }
\label{Fig.2}
 \end{center}
\end{figure}

To illustrate and confirm the possibility of a diode-like quantum gate by applying the non-adiabatic time-asymmetric (NA-TAS) mechanism we apply the above protocol to H$_2^+$ in a laser field. Fig.~\ref{Fig.2}
demonstrates the interaction of H$_2^+$  with a chirped laser pulse according to the two different protocols shown in Fig.~\ref{Fig.1} (two different closed loops  in laser wavelength-intensity parameter plane which encircle the same EP).  We chose this system since it was a subject of experimental
studies for many years (see for example Ref.\cite{NATAN1}) and since our theoretical predictions can be experimentally confirmed.
The numerical results of Fig.\ref{Fig.2} were obtained by propagating the system in the basis of the instantaneous solutions of the H$_2^+$ interacting with the laser.

Note  that in order to design an experiment we need to add the rotational motion into our calculations since the EPs due to the coalescence of different rotational states, rather than different vibrational states, are more likely to happen. The study of the effect of the molecular rotations on the controllable optimal pathway between different ro-vibrational states is currently ongoing an is supported by the results presented in Fig.~\ref{Fig.2}.

The results of Fig.~\ref{Fig.2} can be summarized in Table~1 to illustrate how by going along a loop in one direction we always obtain one of the coalescing states whereas by going in the other direction we will always wind up in the other state. This result is independent of the initial superposition of the two field-free states.  These results clearly demonstrate the power of the mechanism proposed in this research with the purpose to induce asymmetric transition which is caused by non-adiabatic interaction between meta-stable states.
 \begin{table} [!]
\setlength{\tabcolsep}{8pt}
\setlength{\extrarowheight}{3pt}
\begin{centering}
\begin{tabular}{|>{\Huge}c||c|c|c|}
  \hline
     & initial & final (adiabatic) & final (exact)\\
  \hline
  \multirow{2}{*}[1pt] {$\circlearrowright$}
   & $|9\rangle$ & $|10\rangle$ & $|9\rangle$ \\
  \cline{2-4}
  & $|10\rangle$ & $|9\rangle$  & $|9\rangle$ \\ \hline
    \multirow{2}{*}[1pt] {$\circlearrowleft$}
   & $|9\rangle$ & $|10\rangle$ & $|10\rangle$ \\
  \cline{2-4}
  & $|10\rangle$ & $|9\rangle$  & $|10\rangle$ \\ \hline
  \end{tabular}

\caption{ The driven asymmetric effect on the state-to-state vibrational transitions of $H_2^+$ when either a positive or negative chirp laser pulses are used.In a separate column we show that the adiabatic approximation yields incorrect prediction for the final state. In both cases the laser parameters are varied in a closed loop which encircles an EP as shown in Fig.~\ref{Fig.1}.}
  \label{Table1}
\end{centering}
\end{table}

What is missing in this representation is the fact that the NA-TAS mechanism is not only very efficient in producing
asymmetric molecular diode-like device but also increases the quantity of molecules which survive the laser field and have not dissociated. In order to illustrate this important property of the NA-TAS  mechanism we present below the results obtained for many different realizations where the duration of the laser pulse and the maximum field intensity are varied. We examine the effect of the presence of the EP inside  or outside the closed  loop in the chirped laser parameter space on the populations of the propagated wavepacket of $H_2^+$.
The parameter $\rho$ measures the proximity of the EP to the loop and when the EP is inside the loop then $\rho>0$ while when it is outside $\rho<0$.
\begin{figure}[htb]
\begin{center}
\subfloat[][]{
\includegraphics[width=1\columnwidth, angle=0,scale=0.45,
draft=false,clip=true,keepaspectratio=true]{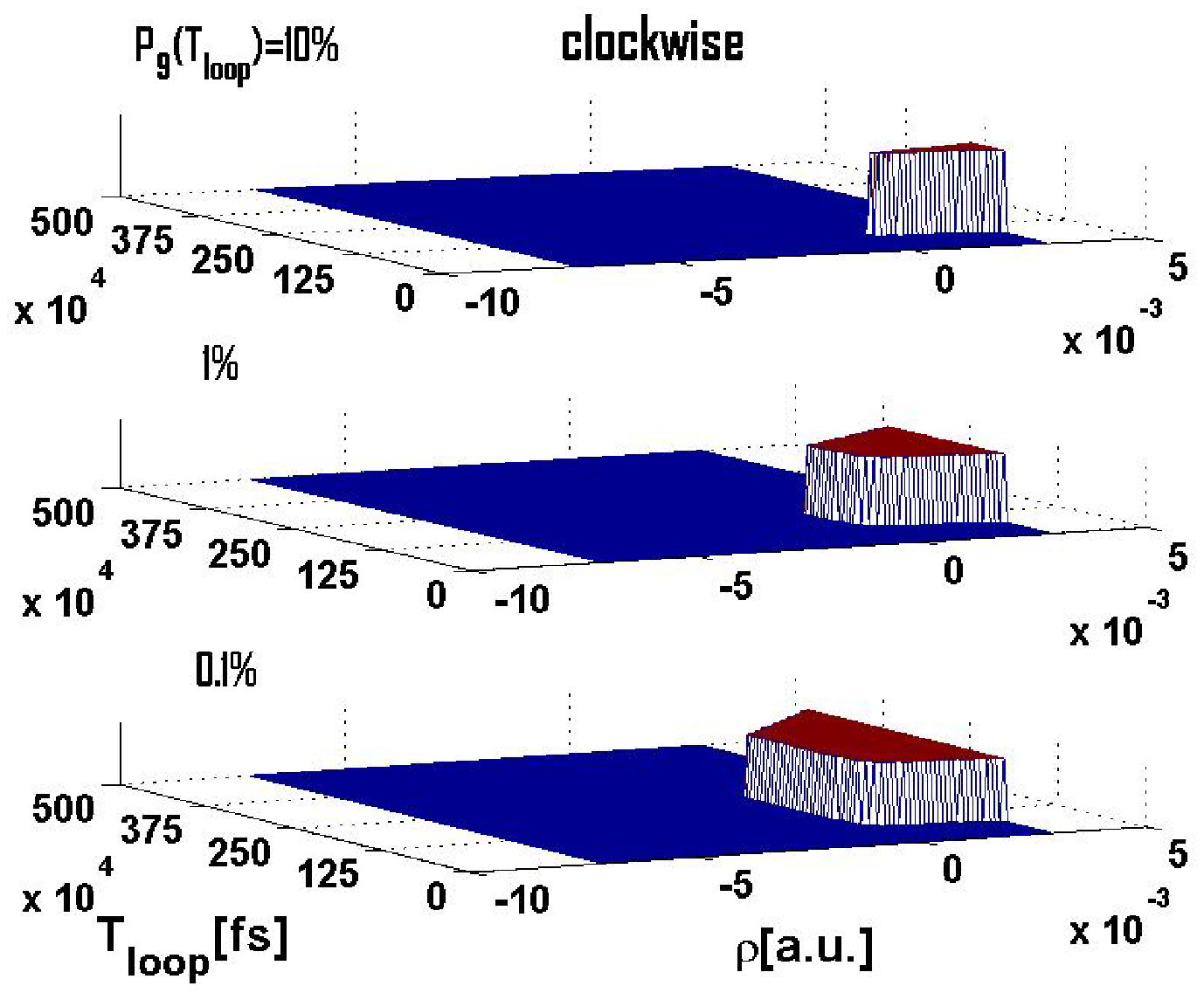}}
\quad
\subfloat[][]{ \includegraphics[width=1\columnwidth, angle=0,scale=0.45,
  draft=false,clip=true,keepaspectratio=true]{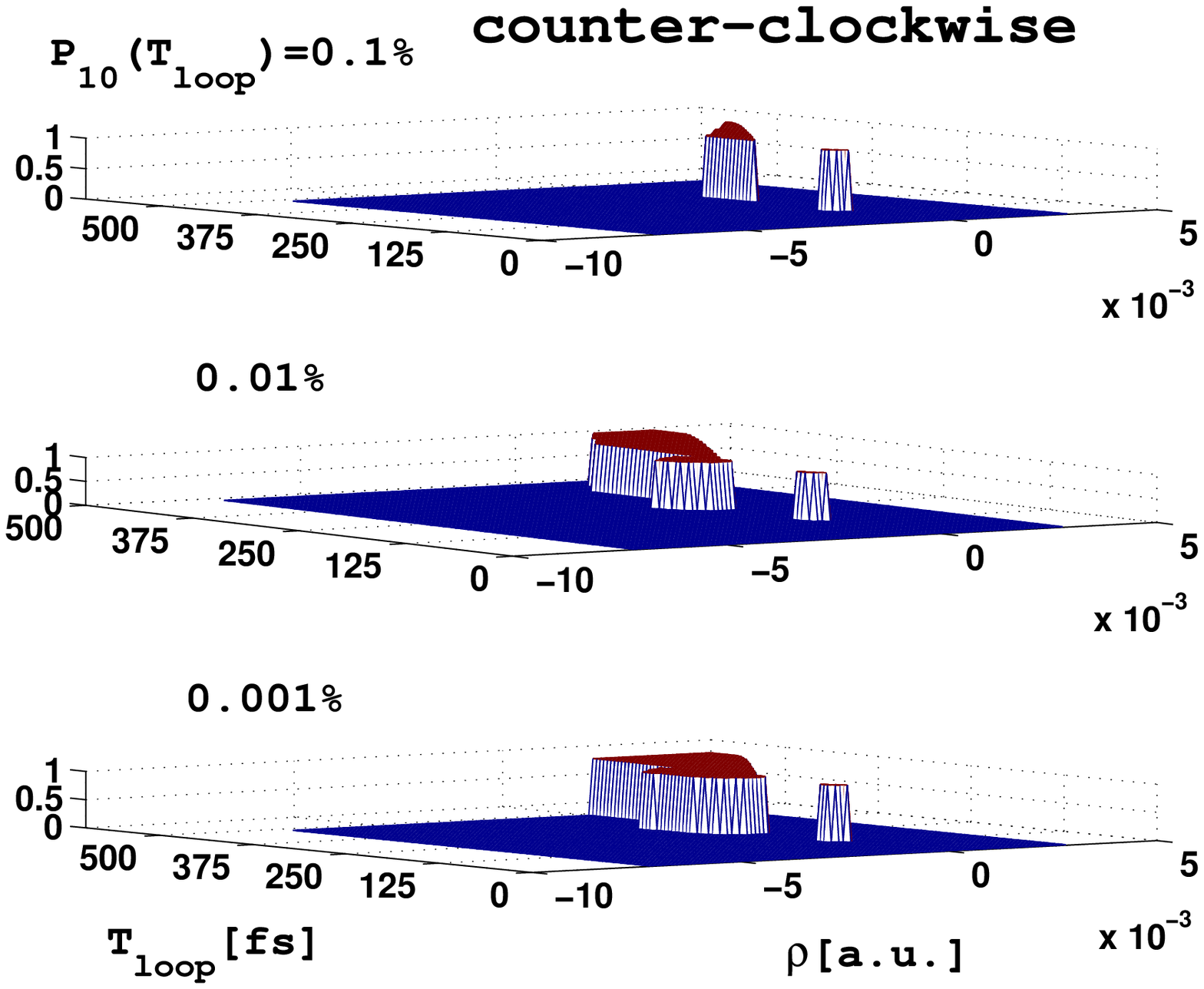}}
\caption{\\(a) The regions in parameter space where the population of the $9th$ vibrational state of $H_2^+$ is at least 1000 larger than the population of the $10th$ vibrational state by the end of the laser pulse which is  positively chirped. Initially at $t=0$ the two vibrational states of $H_2^+$ are equally populated. In the top panel 10$\%$ of the initial population remains, in the middle panel 1$\%$ remains and in the bottom panel 0.1$\%$ survives.\\
(b) The regions in parameter space where the population of the $10th$ vibrational state of $H_2^+$ is at least 1000 larger than the population of the $9th$ vibrational state by the end of the lase pulse which is  positively chirped. Initially at $t=0$ the two vibrational states of $H_2^+$ are equally populated. In the top panel 0.1$\%$ of the initial population remains, in the middle panel 0.01$\%$ remains and in the bottom panel 0.001$\%$ survives.}
\label{PwithS}
\end{center}
\end{figure}

It is evident from the results of Fig.\ref{PwithS} that the relevant regions in parameter space occur for positive $\rho$ values, i.e., when the loop encircles the EP.   This clearly shows the efficiency of the NA-TAS mechanism first proposed here. The application of the NA-TAS protocol results in a diode-like behavior of mixed-to-pure state transitions of $H_2^+$ when the closed loop in the frequency-intensity laser parameter space encircles the EP.
However, above all,  the most significant result presented here is a fundamental one.  It lies in the possibility to observe an
  asymmetric dynamical phenomenon in photo-dissociative experiments that  reflects the effect of an EP
  on the dynamics. This effect is very different from the effect obtained when the degenerate states are bound states
  and the system is hermitian.

Here we showed that there are observable physical phenomena which are hard to explain by hermitian quantum mechanics but can be readily explained and predicted by using the theoretical tools developed within the framework of non-hermitian quantum mechanics. Moreover, the time-asymmetric state-exchange mechanism,  which is based on our ability to locate non-hermitian degeneracies (so called EPs), enables one to control the dynamics  by external parameters of the electromagnetic fields with which the systems under study interact.
This will open a door to new types of technologies, to new type of photo-switches, diode-like atomic, molecular and optical devices. 

\begin{acknowledgments}
ISF (grant 298/11) is acknowledged for their support. A.M. is grateful for the hospitality and support during his stay in Technion.
\end{acknowledgments}

\end{document}